\documentclass[%
 reprint,
 amsmath,amssymb,
 aps,
]{revtex4-2}

\usepackage{graphicx}
\usepackage{dcolumn}
\usepackage{bm}

\begin{document}

\preprint{APS/123-QED}

\title{Net-proton number cumulant ratios as function of beam energy from an expanding nonequilibrium chiral fluid}

\author{Christoph Herold}
 \email{herold@g.sut.ac.th}
\author{Ayut Limphirat}
\author{Poramin Saikham}
\affiliation{
 Center of Excellence in High Energy Physics \& Astrophysics, Suranaree University of Technology, Nakhon Ratchasima, 30000, Thailand}

\author{Marlene Nahrgang}
\affiliation{
SUBATECH UMR 6457, IMT Atlantique, Université de Nantes, IN2P3/CNRS, 4 rue Alfred Kastler, Nantes, 44307, France
}

\author{Tom Reichert}
\author{Marcus Bleicher}
\altaffiliation[Also at ]{GSI Helmholtzzentrum f\"{u}r Schwerionenforschung GmbH, Planckstr. 1, Darmstadt, 64291, Germany}
\affiliation{
Institut f\"{u}r Theoretische Physik, Goethe Universit\"{a}t Frankfurt, Max-von-Laue-Strasse 1, Frankfurt am Main, 60438, Germany
}
\affiliation{
Helmholtz Research Academy Hesse for FAIR (HFHF), GSI Helmholtz Center for Heavy Ion Physics, Campus Frankfurt, Max-von-Laue-Str. 12, Frankfurt, 60438, Germany
}

\date{\today}

\begin{abstract}
The beam energy scan program at RHIC provides data on net-proton number fluctuations with the goal to detect the QCD critical end point and first-order phase transition. Interpreting these experimental signals requires a vital understanding of the interplay of critical phenomena and the nonequilibrium dynamics of the rapidly expanding fireball. We study these aspects with a fluid dynamic expansion coupled to the explicit propagation of the chiral order parameter sigma via a Langevin equation. Assuming a sigma-proton coupling through an effective proton mass, we relate cumulants of the order parameter and the net-proton number at freeze-out and obtain observable cumulant ratios as a function of beam energy. We emphasize the role of the nonequilibrium first-order phase transition where a mixed phase with gradual freeze-out can significantly alter the cumulants. We find that the presence of a critical end point is clearly visible in the cumulant ratios for a relatively wide range of center-of-mass energies. 
\end{abstract}

\maketitle


\section{Introduction}
\label{sec:introduction}

Experiments at SPS and STAR have found that nuclear matter at high temperatures undergoes a transition from a hadronic phase to a quark-gluon plasma \cite{Heinz:2000bk,STAR:2005gfr}, a state of deconfinement and chiral symmetry restoration. While this transition is a continuous crossover for zero and small baryochemical potential $\mu_{\rm B}$ \cite{Aoki:2006we,Borsanyi:2010bp,Bazavov:2014pvz}, a critical endpoint (CEP) and first-order phase transition (FOPT) are expected at large $\mu_{\rm B}$. Here, a variety of effective models of quantum chromodynamics (QCD) \cite{Scavenius:2000qd,Schaefer:2004en,Fukushima:2008wg} as well as functional techniques \cite{Fischer:2014ata,Gao:2020fbl} yield widely different results regarding existence and location of CEP and FOPT in the space of $T$ and $\mu_{\rm B}$. 

Besides these theoretical efforts, considerable workforce is invested in ongoing experimental programs aiming at understanding the QCD phase diagram, e.g. the beam energy scan at STAR \cite{STAR:2021iop}, NA49/61 \cite{Grebieszkow:2009jr,Andronov:2018ccl}, HADES \cite{HADES:2020wpc}, or the upcoming facilities NICA \cite{nica:whitepaper} and FAIR \cite{Friman:2011zz}. Important observables in this context are cumulants of conserved quantities, namely baryon number, strangeness, and electric charge. As shown in lattice QCD \cite{Karsch:2016yzt,Bazavov:2020bjn}, various studies of effective models \cite{Skokov:2010uh,Almasi:2017bhq,Wen:2018nkn}, and functional techniques \cite{Isserstedt:2019pgx}, these cumulants diverge at the CEP and behave characteristically in the critical region around the CEP. Even though these calculations are based on equilibrium thermodynamics, it is widely believed that a measurement of cumulants or cumulant ratios in heavy-ion collisions should reveal nonmonotonic behavior as function of center-of-mass energy in the presence of a CEP. Here, however, a thorough understanding of the nonequilibrium dynamics and a design of proper experimental methods are crucial. For an overview of CEP physics at STAR, see \cite{Luo:2017faz}. 
Since cumulants of various order are directly proportional to certain powers of the correlation length, they diverge in an infinitely large equilibrated medium close to the CEP. In a heavy-ion collision, their growth is limited not only by the finite system size but also by critical slowing down \cite{Berdnikov:1999ph}. The importance of a proper dynamical description of the critical dynamics has been emphasized by a variety of publications in recent years \cite{Nahrgang:2011mv,Herold:2013bi,Mukherjee:2015swa,Jiang:2015hri,Herold:2017day,Stephanov:2017ghc,Stephanov:2017wlw,Nahrgang:2018afz,Nahrgang:2020yxm,Du:2020bxp}. The direct impact on experimental observables, e.g. the net-proton number fluctuations, has been argued and demonstrated in \cite{Athanasiou:2010kw,Stephanov:2011pb,Jiang:2015cnt,Herold:2016uvv}. Equally important than understanding the complicated dynamics near the CEP is understanding and modeling of evolutions passing through a FOPT where spinodal decomposition enforces density inhomogeneities within single events \cite{Randrup:2009gp,Randrup:2010ax,Steinheimer:2012gc,Herold:2013qda,Herold:2014zoa,Jiang:2017fas,Poberezhnyuk:2020cen} and leads to an increased production of entropy \cite{Csernai:1992as,Herold:2018ptm} which could be observed, e.g., via an enhancement of the pion-to-proton ratio.

In this paper, we apply the nonequilibrium chiral fluid dynamics model \cite{Nahrgang:2011mg} to a longitudinal Bjorken expansion along the beam axis \cite{Herold:2018ptm}. This model describes the dynamics of the sigma field as the chiral order parameter with a Langevin equation interacting with a locally thermalized expanding quark fluid. Field and fluid exchange energy-momentum via a source term. We extract cumulants of the sigma field on an event-by-event basis along a parametrized freeze-out curve. As shown in \cite{Stephanov:2011pb}, we relate these cumulants to net-proton number fluctuations by assuming a superposition of standard Poisson and critical fluctuations. We correct our results by the effect of volume fluctuations as detailed in \cite{Skokov:2012ds}. In the present work, we aim at revealing possible signatures in the net-proton number cumulants that would confirm or rule out a CEP and FOPT in stronly-interacting matter. 

After a description of the model in Section~\ref{sec:model}, we report our results on various net-proton number cumulant ratios as function of beam energy in comparison to data from STAR and HADES in Section \ref{sec:results}, and finally conclude with a summary in Section \ref{sec:summary}.

\section{Model description}
\label{sec:model}

The model is based on the Lagrangian of the widely studied quark-meson model \cite{Scavenius:2000qd,Mocsy:2004ab,Schaefer:2004en} with a CEP at $(T_{\rm CEP}, \mu_{\rm CEP})=(100,200)$~MeV. Although arguably simple, the model provides a description of a chiral phase diagram with the generic features of a crossover, CEP, and FOPT. Currently, there is no agreement within the theoretical physics community on where to find the CEP in the phase diagram \cite{Pandav:2022xxx}, yet recent QCD-based calculations from functional renormalization group techniques \cite{Fu:2019hdw,Gao:2020fbl} suggest a location similar to the one in the quark-meson model. We use the fluctuation of the sigma field as the critical mode of the QCD CEP, characterized by a vanishing sigma mass which has also provided one of the fundamental motivations for the beam energy scan program at RHIC \cite{Stephanov:2011pb,Luo:2015ewa}.

Here and in the following, $\mu$ denotes the quark chemical potential, thus $\mu=\mu_{\rm B}/3$. The Lagrangian for light quarks $q=(u,d)$ and the chiral order parameter $\sigma$ with potential $U$ reads
\begin{align}
\label{eq:Lagrangian}
 {\cal L}&=\overline{q}\left(i \gamma^\mu \partial_\mu-g \sigma\right)q + \frac{1}{2}\left(\partial_\mu\sigma\right)^2- U(\sigma)~, \\
 U(\sigma)&=\frac{\lambda^2}{4}\left(\sigma^2-f_{\pi}^2\right)^2-f_{\pi}m_{\pi}^2\sigma +U_0~,    
\end{align}
with standard parameters $f_\pi=93$~MeV, $m_\pi=138$~MeV and $U_0$ such that $U(\sigma)=0$ in the ground state. The value of the pion fields has already been set to its vacuum expectation value of zero. The quark-sigma coupling constant $g$ is fixed requiring that $g\sigma$ equals the nucleon mass of $940$~MeV in vacuum. 

The grand potential $\Omega$ in mean-field approximation resembles that of a Fermi gas of quarks with energies $E=\sqrt{p^2+g^2\sigma^2}$ and is evaluated as
\begin{align}
 \Omega_{q\bar q}=-2N_f N_c T\int \frac{\mathrm d^3 p}{(2\pi)^3} & \left[\log\left(1+\mathrm e^{-\frac{E-\mu}{T}}\right)\right. \\ & \left. +\log\left(1+\mathrm e^{-\frac{E+\mu}{T}}\right)\right]~.
\end{align}
Here, $N_f=2$, $N_c=3$ denote the number of light quark flavors and the number of colors.

\subsection{Equations of motion}

We evolve the zero mode or volume-averaged sigma field defined as $\sigma(\tau)=\frac{1}{V}\int\mathrm d^3 x\sigma(\tau,x)$ using a Langevin equation of motion, 
\begin{equation}
 \label{eq:eom_sigma}
 \ddot\sigma+\left(\frac{D}{\tau}+\eta\right)\dot\sigma+\frac{\delta\Omega}{\delta\sigma}=\xi~,
\end{equation}
neglecting spatial fluctuations. 
We describe the expanding fluid using a Bjorken model, and consequently use proper time $\tau$ rather than coordinate time $t$, starting from an initial thermalization at $\tau_0=1$~fm. Consequently, the dots in Eq.~\eqref{eq:eom_sigma} denote derivatives with respect to $\tau$. For our case of purely longitudinal hydrodynamic flow, we set $D=1$ in the Hubble term. The full and proper nonequilibrium dynamics of sigma is encoded in the dissipation coefficient $\eta$ and the stochastic noise $\xi$ which are related by a dissipation-fluctuation relation \cite{Nahrgang:2011mg}, 
\begin{equation}
\label{eq:dissfluct}
 \langle\xi(t)\xi(t')\rangle=\frac{m_{\sigma}\eta}{V}\coth{\left(\frac{m_{\sigma}}{2T}\right)}\delta(t-t')~.
\end{equation}
Here, $\xi$ is assumed Gaussian and white, i.e. it is not correlated over time. The coefficient $\eta$ includes effects from various processes:
\begin{itemize}
    \item Mesonic interactions, i.e.\  $\sigma\sigma\leftrightarrow\sigma\sigma$ (scattering of a condensed sigma with a thermal sigma), and  $\sigma\leftrightarrow\pi\pi$ (two-pion decay) \cite{Csernai:1999ca}, described by a phenomenological damping coefficient of $\eta=2.2/$fm \cite{Biro:1997va} wherever kinematically allowed. 
    \item Meson-quark interactions, $\sigma\leftrightarrow q\bar q$, leading to a $T$- and $\mu$-dependent coefficient \cite{Nahrgang:2011mg}, 
    \begin{equation}
        \eta=\frac{12 g^2}{\pi}\left[1-2n_{\rm F}\left(\frac{m_\sigma}{2}\right)\right]\frac{1}{m_\sigma^2}\left(\frac{m_\sigma^2}{4}-m_q^2\right)^{3/2}~.
    \end{equation}
\end{itemize}

We assume an ideal fluid of quarks and antiquarks described by the energy-momentum tensor $T^{\mu\nu}_q=(e+p)u^{\mu}u^{\nu}-pg^{\mu\nu}$. Due to energy-momentum conservation the divergence of the total energy-momentum tensor $T^{\mu\nu}_q+T^{\mu\nu}_\sigma$ vanishes, which leads to the evolution equation for the energy density,
\begin{equation}
\label{eq:eom_eden}
 \dot e=-\frac{e+p}{\tau}+\left[\frac{\delta\Omega_{q \bar q}}{\delta\sigma}+\left(\frac{D}{\tau}+\eta\right)\dot\sigma\right]\dot\sigma~,
\end{equation}
while the net-baryon density simply follows
\begin{equation}
 \label{eq:eom_nden}
 \dot n = -\frac{n}{\tau}~.
\end{equation}
In Eqs.~\eqref{eq:eom_sigma} and \eqref{eq:eom_eden}, the pressure is given by $p=-\Omega_{q\bar q}$. It is an explicit function of $\sigma$ which during the evolution is not fixed to its equilibrium value. The fireball volume which appears in Eq.~\eqref{eq:dissfluct} and also later in the description of the freeze-out cumulants, is given by $V=\pi R^2 \tau$, with a gold nucleus radius of $R=7.3$~fm, assuming central Au+Au collisions.

\subsection{Freeze-out and mapping to beam energies}

We investigate higher order cumulant ratios of the net-proton number along a freeze-out curve which has been obtained from thermal model fits to experimental data from SIS, AGS, SPS, and RHIC for a wide range of beam energies from $\sqrt{s}=2.24$ to $200$~AGeV \cite{Cleymans:2005xv}. The parametrization reads
\begin{equation}
    \label{eq:freeze}
    T_{\rm f.o.}(\mu_B) = a - b\mu_B^2 - c\mu_B^4~,
\end{equation}
with $a=0.166$~GeV, $b=0.139$~GeV$^{-1}$ and  $c=0.053$~GeV$^{-3}$. Since the phase boundary of the quark-meson model would lie below the thus obtained freeze-out line, we scale $T$ and $\mu$ in this parametrization with a common factor $T_{\rm crossover}(\mu=0)/T_{\rm freeze}(\mu=0)$ yielding a freeze-out curve which coincides with the crossover line at $\mu=0$ and consistently lies below the crossover and phase transition for $\mu>0$. Here, $T_{\rm crossover}(\mu=0)=145$~MeV is determined by a maximum of the quark number susceptibility.  

We define initial conditions for the evolution by adopting initial values $T_{\rm i}$ and $\mu_{\rm i}$ from previous studies \cite{Herold:2018ptm}. During the evolution of the fluid according to Eqs.~\eqref{eq:eom_sigma}, \eqref{eq:eom_eden}, \eqref{eq:eom_nden}, the trajectory in $T$-$\mu$ space will eventually hit the freeze-out curve. This initial hit point is then used to map the evolution to a corresponding beam energy via:
    \begin{equation}
    \label{eq:matchsqrts}
    \mu_B(\sqrt{s}) = \frac{d}{1 + e\sqrt{s}}~,
    \end{equation}
with parameters $d = 1.308$ GeV and  $e = 0.273$ GeV$^{-1}$ determined in accordance with the freeze-out curve above \cite{Cleymans:2005xv}. The baryochemical potential in Eq.~\eqref{eq:matchsqrts} is obtained from averaging over events with the same initial condition. Note that the thus obtained beam energies are to be understood as guidelines to put our results into the context of experimentally obtained data from STAR and HADES. Even though in our present model, the CEP is passed for evolutions with an initial $\sqrt{s}\sim 5$~GeV, this does not mean that we necessarily expect the physical CEP to be found there. However, it puts us in a position to estimate a CEP's impact on measurable observables in a fully dynamical nonequilibrium setup if it indeed exists in this low-energy range.

\subsection{Sigma and net-proton number cumulants}

To relate the fluctuations in the chiral order parameter $\sigma$ to fluctuations or cumulants of the net-proton number, we follow the strategy outlined in \cite{Stephanov:2011pb}. Consider an infinitesimal change of the chiral field, $\delta\sigma$, leading to a change of the effective proton mass by $\delta m = g \delta \sigma$. Assuming a sigma-proton coupling $g\sigma \bar p p$, we may write fluctuations of the momentum space distribution function for protons, $f_k$, as 
\begin{equation}
  \delta f_k = \delta f_k^0 + \frac{\partial n_{\rm FD}}{\partial m}\, g\,\delta\sigma~.
\end{equation}
The first term $\delta f_k^0$ is the purely statistical fluctuation, and in the second term, $n_{\rm FD}$ denotes the Fermi-Dirac distribution for a particle of a given mass $m$. The fluctuation of the net-proton multiplicity $N=V\,d\int \frac{\mathrm d^3 k}{(2\pi)^3} f_k$ is then given by  
\begin{equation}
  \delta N = \delta N^0 + V\, g\, \delta\sigma\,d
\int\frac{\mathrm d^3 k}{(2\pi)^3} \frac{\partial n_{\rm FD}}{\partial m}\,,
\end{equation}
where $d=2$ is the spin degeneracy factor. The first term $\delta N^0$ can be assumed Poisson distributed. Consequently, all of its cumulants are equal to the expectation value $\langle N\rangle$. In leading order and assuming no correlations between $\delta N^0$ and $\delta\sigma$, we can express cumulants of order $n$ as
\begin{equation}
  \label{eq:N4}
  \langle\delta N^n\rangle_c = \langle N\rangle + \langle \delta\sigma_V^n\rangle_c
\left( g\,d\int\frac{\mathrm d^3 k}{(2\pi)^3}\frac{\partial n_{\rm FD}}{\partial m}\right)^n~.
\end{equation}

In this notation, $\sigma_V=\int \mathrm d^3 x \sigma = \sigma V$ as we neglect spatial fluctuations, and $\langle \cdot \rangle_c$ is the respective cumulant, which is equal to the expectation value for $n=1$ and to the corresponding central moment for $n= 2, 3$. For $n=4$, we have 
\begin{equation}
    \langle \delta\sigma_V^4\rangle_c=\langle \delta\sigma_V^4\rangle - 3\langle\delta\sigma_V^2\rangle^2~,
\end{equation}
and similarly for $\langle\delta N^n\rangle_c$. In the next section~\ref{sec:results}, we will use the shorter notation $C_n$ for the net-proton number cumulants $\langle\delta N^n\rangle_c$ which has also been commonly used in experimental studies of recent years.

\section{Research procedure and results}
\label{sec:results}

We initialize the fluid at a set of fixed values ($T_i$,$\mu_i$) and define initial sigma field, energy density, quark number density, and pressure as the corresponding equilibrium values. The pairs of initial values are hereby adopted form earlier works \cite{Herold:2018ptm,Herold:2022tki} and will be matched to center-of-mass energies using the freeze-out condition, Eq.~\eqref{eq:matchsqrts}. The coupled system evolves according to the equations of motion until the freeze-out curve is hit. Notably, for expansions at high baryochemical potential, the freeze-out curve can be hit more than once due to the nonequilibrium evolution of the expanding plasma. This effects occurs due to the sudden release in latent heat that drives the system back into the chirally symmetric phase, visible in a back-bending of the trajectories in $T$ and $\mu$ \cite{Herold:2013qda,Herold:2018ptm}. In contrast to that, an equilibrated hydrodynamic system with constant $S/A$ would pass along the phase boundary for a finite amount of time \cite{Steinheimer:2007iy}. To take into account this effect of a possible mixed phase in a heavy-ion collision and its impact on the observed cumulants, we evolve the system for these cases until a second crossing of the freeze-out curve and subsequently calculate cumulants at both crossing or hit points. In the following figures and text, we consequently refer to the "first hit" as the cumulants evaluated at the first crossing of the freeze-out line and the "second hit" as those evaluated at the second crossing, where applicable. This is relevant for evolutions near the CEP in the phase diagram or crossing the FOPT line \cite{Herold:2022tki}. Ultimately, it will provide us with a range of possible cumulants for the respective energies. The characteristic back-bending after passing the phase boundary becomes most pronounced for evolutions passing the FOPT due to significant energy dissipation pushing the system back into the chirally restored phase. Physically, this process could manifest in a gradual freeze-out where the medium undergoes droplet formation \cite{Randrup:2009gp,Mishustin:1998eq}. Another suggested signal for such a delayed transition process is an enhancement in the dilepton production~\cite{Seck:2020qbx}. 

We simulate $10^7$ events and calculate event-by-event fluctuations in terms of cumulants of $\sigma_V$. Since these are subject to significant fluctuations of the freeze-out volume, we include the corresponding corrections as derived in \cite{Skokov:2012ds}. This is most relevant for evolutions at the lowest energies with large variations in the time at which different events from the same initial condition hit the freeze-out curve \cite{Herold:2022tki}. Eq.~\eqref{eq:N4} allows us to determine the net-proton number cumulants. The such obtained values are compared with a Poisson baseline. For the net-proton number ($p-\bar p)$, the cumulants assuming Poisson distributed proton and antiproton numbers are calculated by
\begin{equation}
    \label{eq:poisson}
    C_{n, p-\bar p}=C_{n, p}+(-1)^n C_{n, \bar p}~,
\end{equation}
where $C_{n, p}$, $C_{n, \bar p}$ are equal to the expectation values of the Poisson distribution for all orders $n$, cf.~\cite{STAR:2021iop}. We furthermore provide comparison to the equilibrium net-baryon number susceptibilities which are obtained as
\begin{equation}
    \label{eq:susc}
    \chi_n = \frac{\partial^n (p/T^4)}{\partial(\mu_{\rm B}/T)^n}~.
\end{equation}

\begin{figure}[tbp]
\centering
\includegraphics[width=.49\textwidth,origin=c]{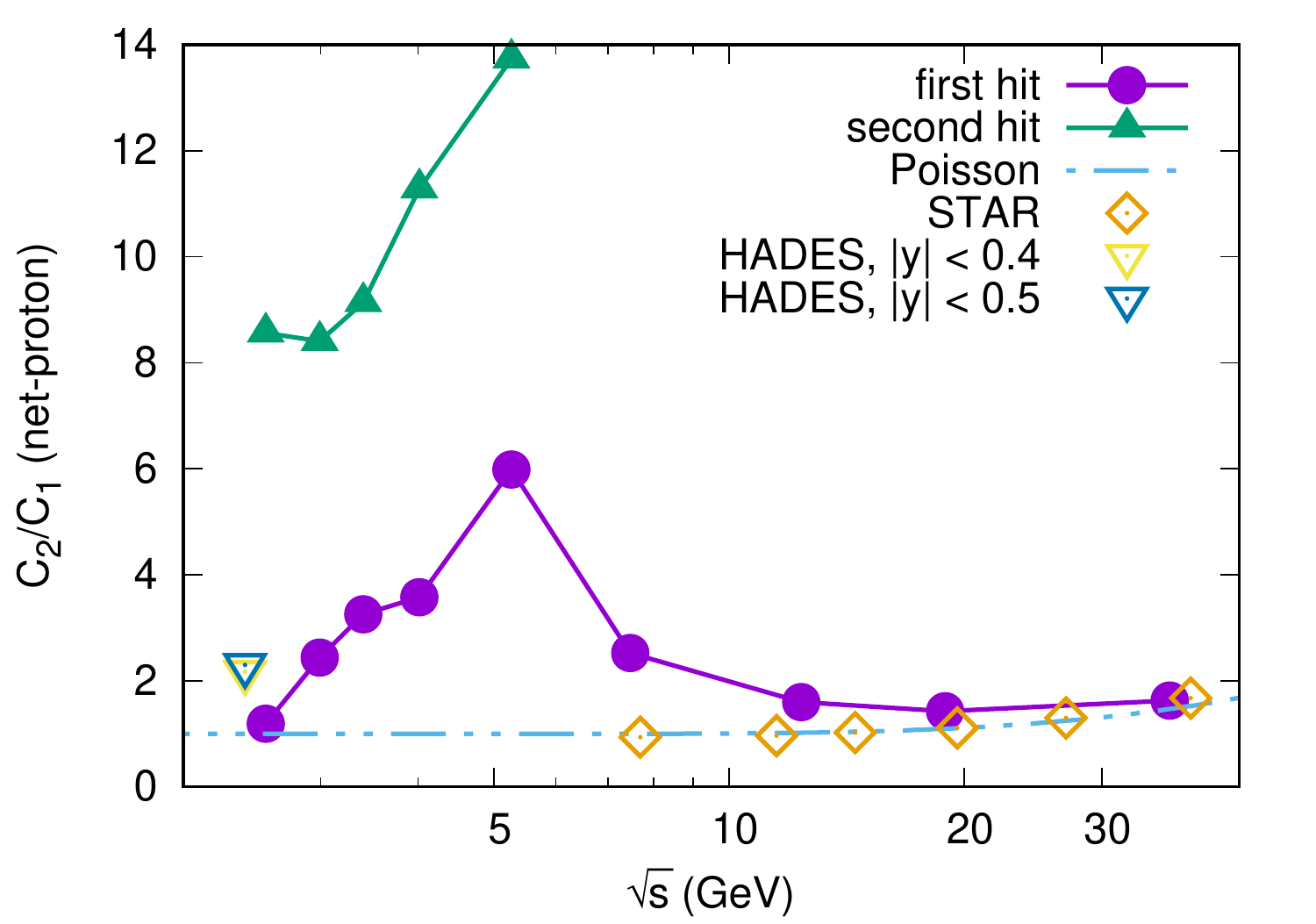}
\hfill
\includegraphics[width=.49\textwidth,origin=c]{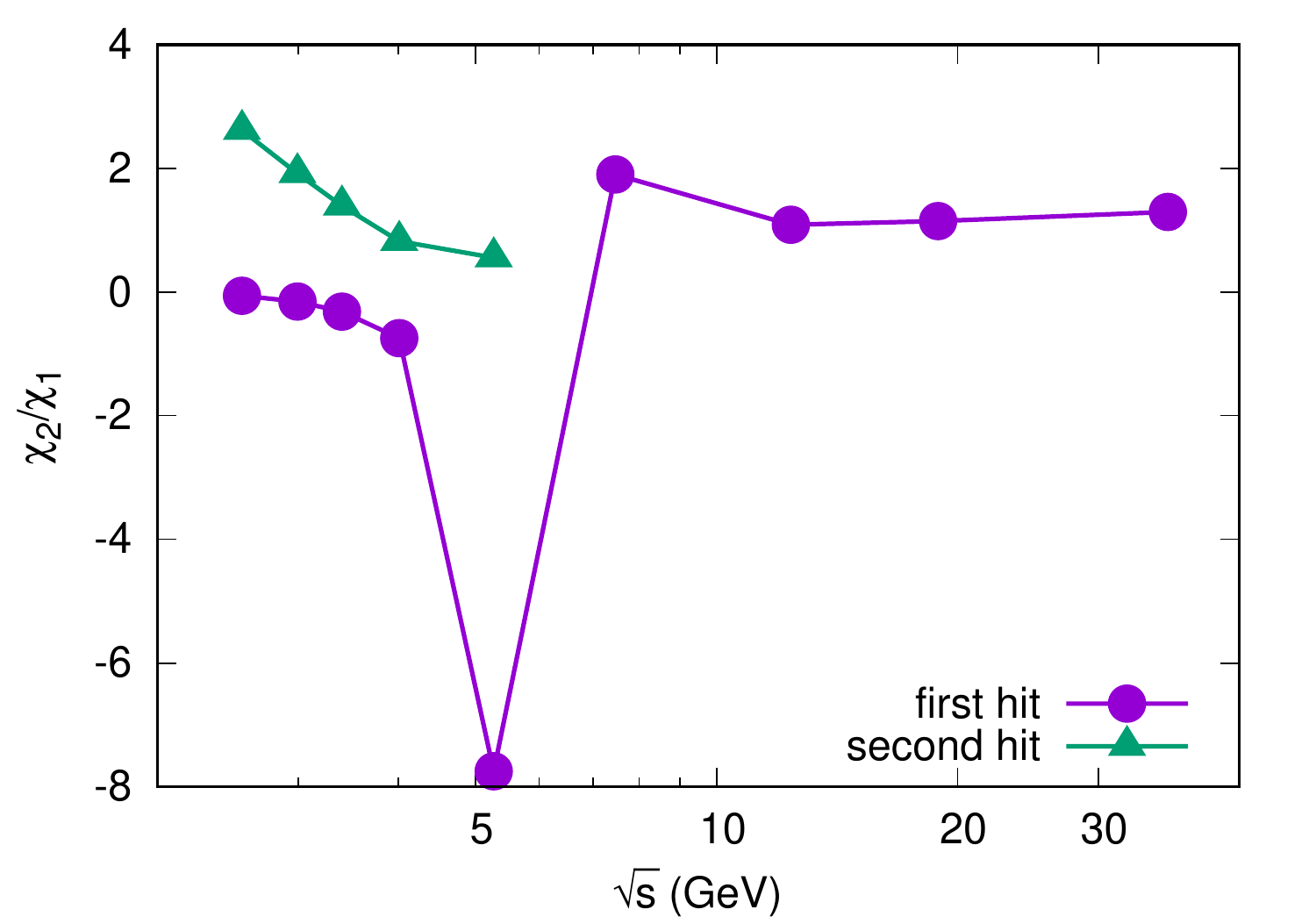}
\caption{Cumulant ratios of the net-proton number are enhanced around the CEP (top). Susceptibility ratios for comparison (bottom).}
\label{fig:c2c1}
\end{figure}

Figure~\ref{fig:c2c1} (top) shows the ratio $C_2/C_1$ of the net-proton number compared to results from STAR for energies $\sqrt{s_{\rm NN}}\ge 7.7$~GeV \cite{STAR:2021iop} and HADES for $\sqrt{s_{\rm NN}}= 2.4$~GeV \cite{HADES:2020wpc}. Since the HADES collaboration reported a strong dependence of cumulant ratios on the chosen rapidity window, we depict results for both $\lvert y\rvert<0.4$ and $\lvert y\rvert<0.5$ for comparison, the latter one also being applied to STAR data. In \cite{HADES:2020wpc}, a reliable rapidity window of $\lvert y\rvert<0.46$ is quoted. We see that for high energies, our results together with STAR data lie close to the Poisson baseline. With decreasing energy, our model yields values that are gradually enhanced until they reach a maximum at around $5$~GeV for the evolution that passes through or close by the CEP and freezes out close to a spinodal line of the FOPT \cite{Herold:2022tki}. Along the spinodal lines, all susceptibilities diverge in nonequilibrium with critical exponents that become larger with order of the susceptibility \cite{Sasaki:2007qh,Herold:2014zoa}. This is clearly reflected in the peak of the corresponding susceptibility ratio visible on the bottom of figure~\ref{fig:c2c1}. Further lowering the energy results in a gradual return to the baseline for the first hit of the freeze-out curve, while for the second hit, significantly larger fluctuations are observed, possibly due to an enhancement of spinodal instabilities. Notably, the data from HADES lies within the thus obtained range of cumulant ratios at the lowest beam energy. Obviously, a significant gap in the experimental data will have to be filled by future experiments such as FAIR and NICA that aim at exploring the high-$\mu_{\rm B}$ region of the QCD phase diagram. The susceptibility ratio on the right hand side of figure \ref{fig:c2c1} shows a similar trend for high energies. The freeze-out close to the spinodal line, where susceptibilities in the presence of spinodal instabilities diverge and change sign \cite{Sasaki:2007qh,Herold:2014zoa}, results in the strongly negative value of $\chi_2/\chi_1$ which is reflected in the peak of $C_2/C_1$, similar to what we found for the other cumulant ratios as we will discuss below. 

\begin{figure}[tbp]
\centering
\includegraphics[width=.49\textwidth,origin=c]{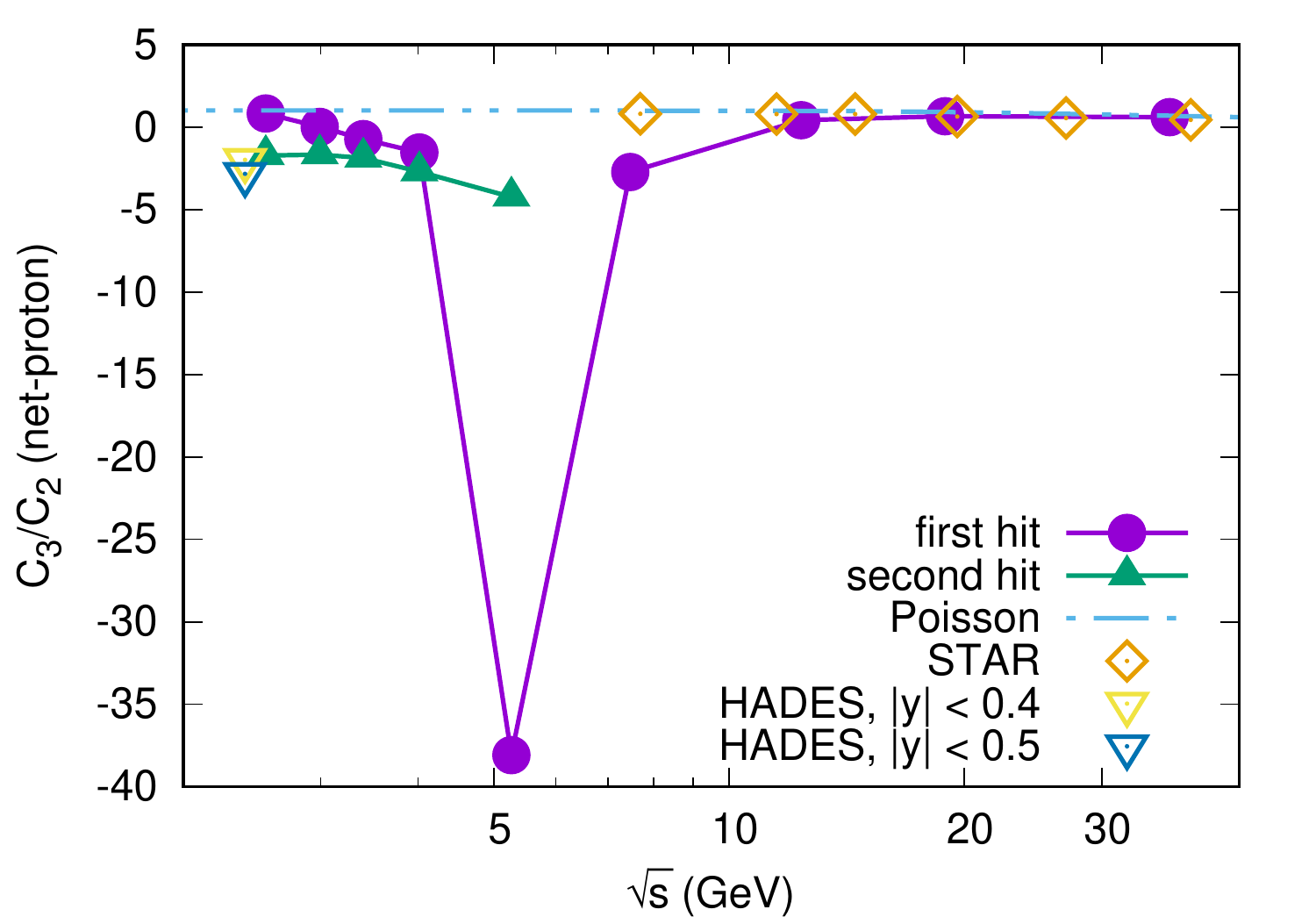}
\hfill
\includegraphics[width=.49\textwidth,origin=c]{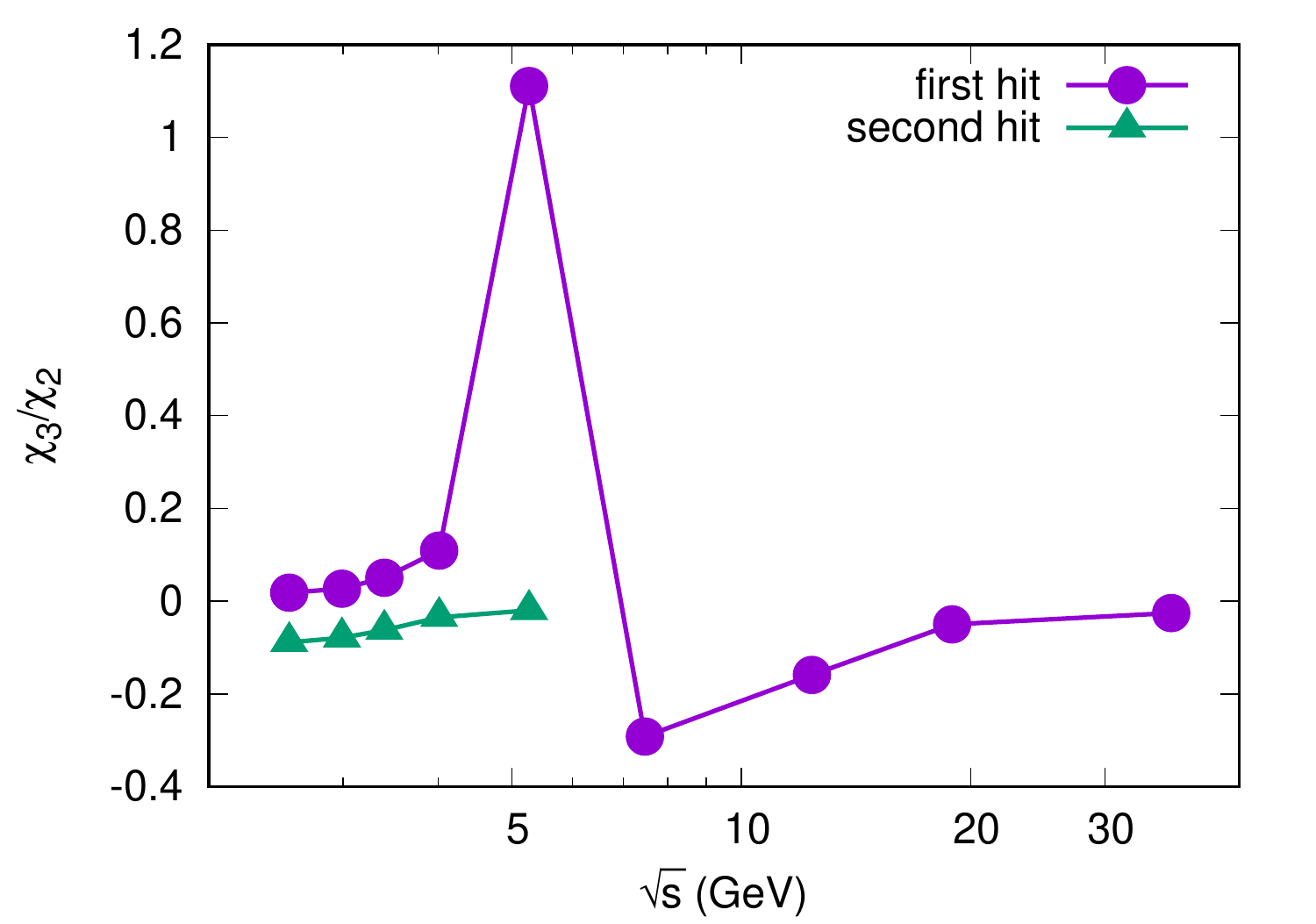}
\caption{Cumulant ratios of the net-proton number are enhanced around the CEP (top). Susceptibility ratios for comparison (bottom).}
\label{fig:c3c2}
\end{figure}

\begin{figure}[tbp]
\centering
\includegraphics[width=.49\textwidth,origin=c]{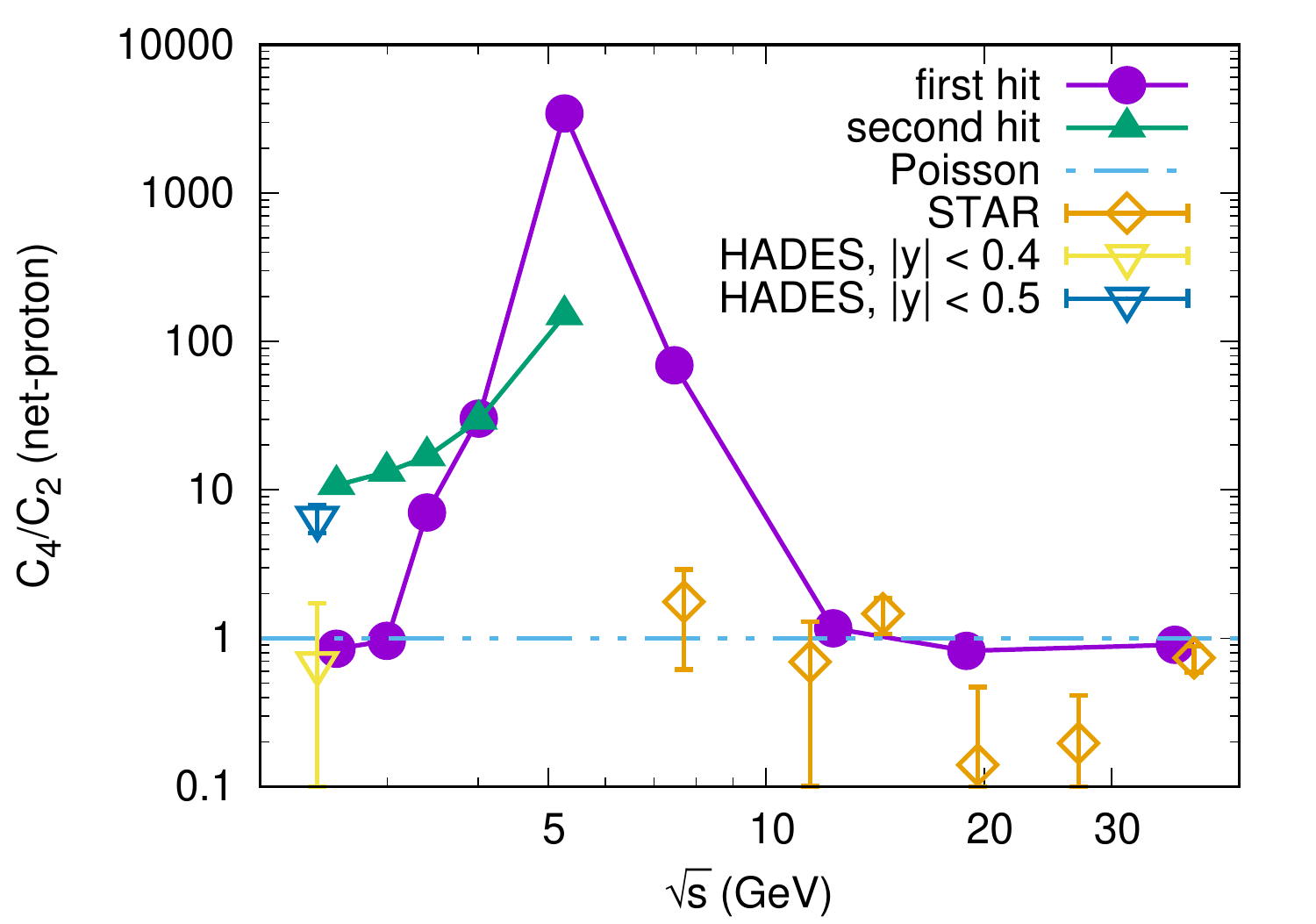}
\hfill
\includegraphics[width=.49\textwidth,origin=c]{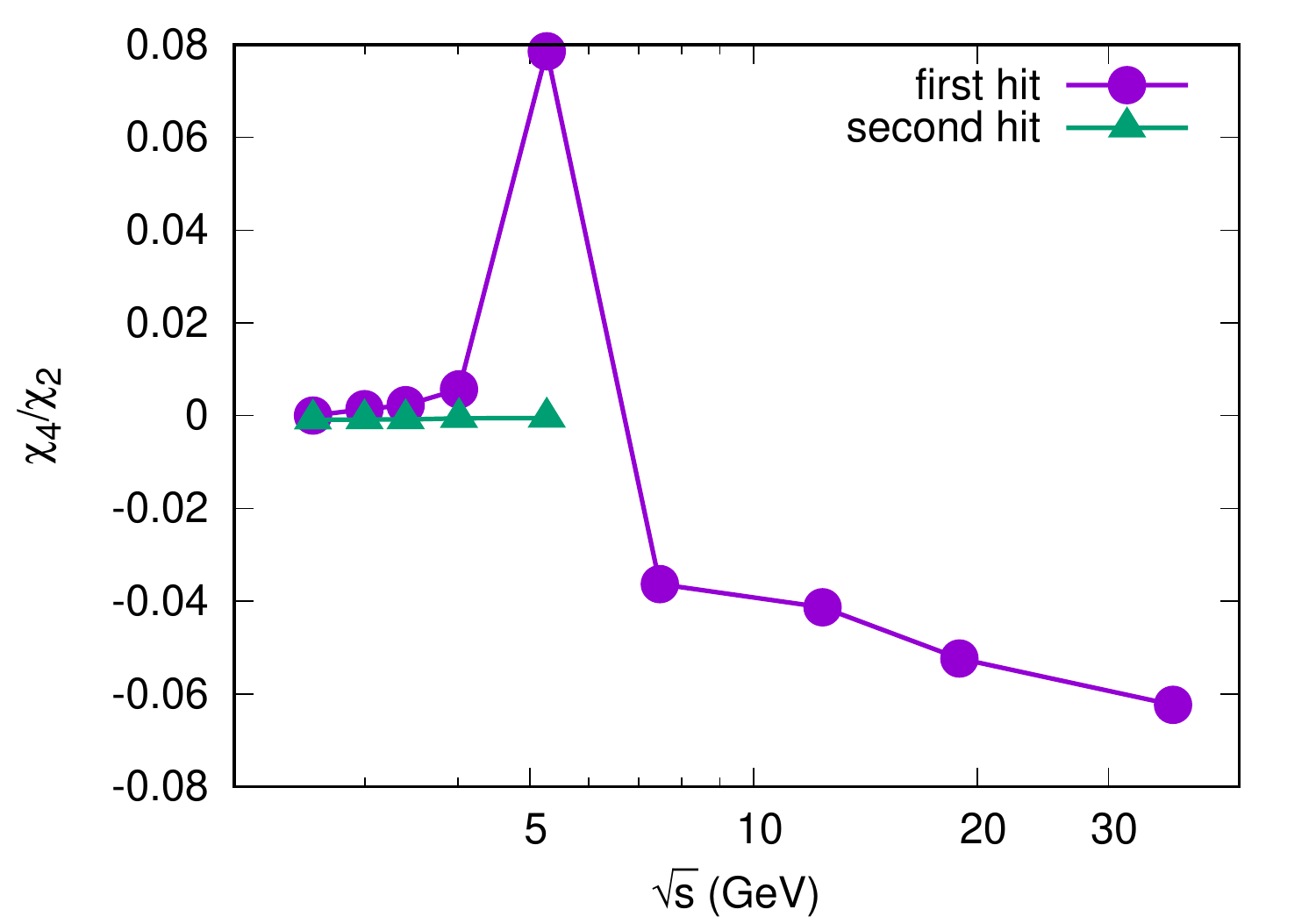}
\caption{Cumulant ratios of the net-proton number are enhanced around the CEP (top). Susceptibility ratios for comparison (bottom).}
\label{fig:c4c2}
\end{figure}

The cumulant ratio $C_3/C_2$ is shown and compared to experimental data in figure~\ref{fig:c3c2} (top). The most notable feature is, again, the strong impact of the CEP around $\sqrt{s}=5$~GeV. Besides that, the obtained points from our model are close to the baseline for high energies and the second hit at the lowest energy is close to the data point from HADES, where a suppression of $C_3/C_2$ was found, possibly a result of the dynamics at the FOPT. This is also found in the net-baryon number susceptibility ratio on bottom part of the same figure, approaching zero for the first hit and remaining negative for the second hit of the parametrized freeze-out curve. The most apparent difference is the sign at the CEP evolution which is positive for the susceptibilities, but negative for the cumulants. As mentioned before, the evolution passing close to the CEP freezes out very close to the spinodal line where some of the susceptibilities change sign \cite{Herold:2014zoa}. Therefore, finite-time effects can here dramatically influence the final values at freeze-out. 

Finally, the ratio $C_4/C_2$ is depicted in figure~\ref{fig:c4c2}, on the top we see qualitative similarities between our model results and the experimental data. A slight suppression in the STAR data around $20$~GeV is also present in our model calculations where the cumulant ratio lies below the Poisson baseline, although with smaller significance. Then, as the beam energy is lowered, the notable point at $7.7$~GeV where $C_4/C_2$ is enhanced is reflected in an enhancement, albeit orders of magnitude larger, of the ratio from our calculation. Here, it is necessary to emphasize that the aforementioned freezing out near the spinodal line leads to larger and larger cumulants at higher orders. Lowering the beam energy even further, our calculations approach the HADES results which for this cumulant ratio show the strongest dependence on the applied experimental cut. Within error bars, both points lie within our range defined by the first and second hit of $C_4/C_2\sim 1-10$. The susceptibility ratios, shown on the bottom of the same figure, are close to zero at these low energies which could indicate that an enhancement of the net-proton number cumulants occurs through a prolonged evolution in the mixed-phase region for the FOPT. The positive peak for the CEP evolution is also found in the susceptibilities, however, for larger energies, the values are slightly negative, which is only partly reflected in the obtained cumulants.

\section{Summary and conclusions}
\label{sec:summary}
We have studied cumulant ratios $C_2/C_1$, $C_3/C_2$, $C_4/C_2$ of the net-proton number at STAR and HADES energies within a nonequilibrium chiral Bjorken expansion. Here, a sigma model served as input for a generic chiral phase structure and net-proton cumulants have been calculated event-by-event from cumulants of the sigma field at a parametrized freeze-out curve. Volume fluctuations have been accounted for and were properly corrected. Although admittedly crude and neglecting effects of an inhomogeneous medium, the dynamical description nevertheless shows some qualitative resemblance to the experimental data, in the approach of the Poisson baseline for high energies far away from the critical region, but also the enhancement or suppression of certain cumulant ratios at a speculated CEP or FOPT. We have demonstrated the general impact of a CEP and FOPT on cumulant ratios as key observables for probing the QCD phase structure. If indeed a CEP is present for center-of-mass energies below $7.7$~GeV, it should be clearly visible in a relatively wide range energy range and manifest itself and the adjacent FOPT through an enhancement and/or suppression of cumulant ratios. Clearly, the current gap in beam energies from $2.4$ for $7.7$~GeV requires filling by future experiments.

Possible future improvements of our current model include the consideration of a spatially inhomogeneous fluid and an extension of the study to full (3+1) dimensional hydrodynamics. 

\begin{acknowledgments}
This work was supported by (i) Suranaree University of Technology (SUT), (ii) Thailand Science Research and Innovation (TSRI), and (iii) National Science Research and Innovation Fund (NSRF), project no. 160355. This research has received funding support from the NSRF via the Program Management Unit for Human Resources \& Institutional Development, Research and Innovation [grant number B16F640076].
\end{acknowledgments}

\bibliographystyle{apsrev4-1}
\bibliography{mybib}

\end{document}